\documentclass[]{article}

\usepackage{graphicx}
\usepackage{amsmath}
\usepackage{authblk}
\usepackage{subfigure}
\usepackage{tikz}
\title{Reconciling cooperation, biodiversity and stability in complex ecological communities}

\author[a]{Chengyi Tu} 
\author[a]{Samir Suweis \thanks{Corresponding author: suweis@pd.infn.it}}
\author[b]{Jacopo Grilli}
\author[c]{Marco Formentin \thanks{Corresponding author: marco.formentin@unipd.it}}
\author[a]{Amos Maritan \thanks{Corresponding author: maritan@pd.infn.it}}
\affil[a]{Department of Physics and Astronomy ``Galielo Galilei'', University of Padova, Via Marzolo 8, 35131 Padova, Italy}
\affil[b]{Department of Ecology and Evolution, University of Chicago, 1101 E. 57th, Chicago, IL 60637, USA}
\affil[c]{Department of Mathematics ``Tullio Levi-Civita'', University of Padova, Via Trieste 63, 35129 Padova, Italy}

\begin{document}

\maketitle

\begin{abstract}
Empirical observations show that ecological communities can have a huge number of coexisting species, also with few or limited number of resources. These ecosystems are characterized by multiple type of interactions, in particular displaying cooperative behaviors. 
However, standard modeling of population dynamics based on Lotka-Volterra type of equations predicts that ecosystem stability should decrease as the number of species in the community increases and that cooperative systems are less stable than communities with only competitive and/or exploitative interactions. 
Here we propose a stochastic model of population dynamics, which includes exploitative interactions as well as cooperative interactions induced by cross-feeding. The model is exactly solved and we obtain results for relevant macro-ecological patterns, such as species abundance distributions and correlation functions. In the large system size limit, any number of species can coexist for a very general class of interaction networks and stability increases as the number of species grows. For pure mutualistic/commensalistic interactions we determine the topological properties of the network that guarantee species coexistence. We also show that the stationary state is globally stable and that inferring species interactions through species abundance correlation analysis may be misleading. Our theoretical approach thus show that appropriate models of cooperation naturally leads to a solution of the long-standing question about complexity-stability paradox and on how highly biodiverse communities can coexist.
\end{abstract}

\section{Introduction}

Research in population dynamics has a long history dating back to almost one thousand year ago with Fibonacci modeling of rabbits population. 
Nevertheless it is still under debate which are the mechanisms allowing the coexistence of many interacting species in the same environment \cite{hutchinson1961paradox, may1972,mccann1998weak, Hubbell2001, levin2014public, coyte2015ecology,posfai2017metabolic}. 
The current loss of earth biodiversity \cite{hull2015rarity} makes this open question of great relevance today more than ever, and this challenge calls for interdisciplinary approaches \cite{fort2013statistical, azaele2016statistical}.
Historically, the Lotka and Volterra (LV) equations \cite{lotka1925, volterra1926} have provided much theoretical guidance and several microscopic derivation of these equations have been proposed \cite{matsuda1992statistical, mckane2004stochastic, constable2015models}. 
Furthermore, these equations are the core of most of the multi-species deterministic population dynamics models based on the ecological concept of niche partitioning: competing species in order to coexist need to interact with the environment differently and to rely on not-overlapping resources \cite{hutchinson1961paradox, may1972}.

While prey-predator and competitive interactions have been extensively studied \cite{goh1977feasibility, hastings1988food, mckane2004stochastic, barabas2012continuous}, mutualistic/commensalistic interactions, which are beneficial to one or both the involved species, have historically received less attention. The current approach to mutualistic population dynamics is a mere generalization of the LV types of models, which does not change the functional form of the two-species interaction phenomenological equations, but utilizes beneficial (+ +) instead of predator-prey (+ -, here called exploitative) interactions \cite{holling1959some, may1972, GarciaAlgarra2014, grilli2017higher, levine2017beyond}. 
In particular, a microscopic derivations of the phenomenological equations specific for the population dynamics in mutualistic communities is still missing. 
Moreover, a generalization of the stability-complexity theorem \cite{may1972, mccann2000diversity} has revealed that mutualism is even more detrimental to stability as the product $SC$ increases \cite{allesina2012stability, suweis2013emergence, suweis2015effect, coyte2015ecology}, where $S$ is the number of species and $C$, the connectivity, is the fraction of non-zero pairwise interactions between species. 
This prediction clashes with the observation of widespread mutualistic interactions (or other facilitating interactions) in many natural communities where the biodiversity is very high \cite{bascompte2007plant,morris2013microbial, wingreen2006cooperation, zelezniak2015metabolic,posfai2017metabolic}, although other cases have been also observed \cite{foster2012competition}.

An alternative theoretical approach to niche-based multi-species deterministic modeling is the Neutral Theory (NT) of Biodiversity \cite{Hubbell2001, volkov2003neutral, alonso2006merits, pigolotti2009speciation, bertuzzo2011spatial, azaele2016statistical, houchmandzadeh2017neutral}. 
In NT organisms of a community have identical per-capita probabilities of giving birth, dying, migrating, and speciating, regardless of the species they belong to. 
In this sense NT is symmetric and aims to model only species on the same trophic level-species therefore competing for the same pool of resources. 
An important example of neutral model is the voter model (VM) \cite{liggett2013stochastic, durrett1999stochastic, sood2005voter, azaele2016statistical}. 
The VM is a paradigmatic model to describe competition in many fields going from social sciences \cite{castellano2009statistical} to biology \cite{pinto2011quasi}. 
In the ecological context one deals with a community of $N$ individuals belonging to $S$ different species. 
In its simplest version, at every time step a randomly selected individual dies and the corresponding resources are freed up for colonization.
An important limitation of this modeling is that it does not explicitly consider  species interactions (e.g. mutualism/commensalism). 

Although it has been already shown that niche based and neutral approaches are only apparently contrasting \cite{segura2010emergent, fisher2014transition}, two crucial issues in the current literature are: (i) the lack of a general framework specifically developed to model mutualistic and commensalistic interactions where species interactions are added on neutral models and can modify birth-death rates; and (ii) understand the role of mutualistic/commensalistic interactions in determining species coexistence and how they impact on patterns such as species abundance distribution \cite{fort2013statistical, azaele2016statistical}. In particular, we propose how to incorporate mutualistic interactions induced by cross feeding \cite{goldford2017emergent,morris2013microbial,pacheco2018costless} in order to have effective equations where resources are not explicitly modelled. We will show that these equations are different from the Lotka-Volterra types typically used until now.

In this work we thus present a theoretical framework where, starting from a VM-like microscopic stochastic modeling, we add interactions among species and we properly account the effect of cooperation and exploitation. These interactions affect neutrality and lead, in their mean field formulation, to an emergent multi species-mutualistic model. 
Reconciling apparently contrasting observations and previous results \cite{may1972, mccann2000diversity, allesina2012stability, suweis2013emergence, coyte2015ecology}, we show that in our model ecosystem cooperation promotes biodiversity and diversity increases its stability.

\section{Results}

\subsection{Cooperative Voter Model with Mutualistic Interactions}

In details, be $\eta_z$ the species label at spatial position $z$, where $\eta_z\in\{1,\ldots,S\}$ and $z=1, \dots, N$. 
The state at time $t$ of the system is given by $\eta(t) = (\eta_1(t),\eta_2(t),\ldots,\eta_{N}(t)) \in \{1,\ldots,S\}^N$. 
We also set $\bar{\eta}^k$ to be the fraction of individuals of the $k$-species. 
We now introduce a directed graph on the set $\{1,\ldots,S\}$, where the nodes correspond to species and directed links represent the network of ecological interactions. 
Such a graph is defined through two matrices $M_{ij}$ (cooperation matrix) and $L_{ij}$ (exploitation matrix) satisfying the following conditions: (i) For all $i,j=1,\ldots,S$, $M_{ij}\geq 0$; (ii) For all $i,j=1,\ldots,S$, it must be $L_{ij} L_{ji}<0$ or $L_{ij}=L_{ji}=0$; (iii) For all $i,j=1,\ldots,S$, we have $L_{ij}M_{ij}=0$, i.e. species $i$ and $j$ can not simultaneously have both mutualistic and exploitative interactions. 
Both intra and inter-species competition is indirectly accounted by fixing the total number of individuals in the community \cite{Hubbell2001, azaele2016statistical}. 

In ecological terms, given two species $i$ and $j$, a directed link of strength $M_{ij}$ from $i$ to $j$ means that the $j$-th species receives a beneficial effect from the interaction with the $i$-th species, while $L_{kl}>0 (<0)$ and $L_{lk}<0 (>0)$ denotes that the $l$-th species exploits (is exploited by) the $k$-th species. 
For instance, in the former case we can think a microbial community where the presence of a certain species creates an environment  for instance by secreting metabolites (cross-feeding), which modifies the niches and favors the growth of other bacteria \cite{zelezniak2015metabolic, goldford2017emergent,pacheco2018costless}; in the latter one, we may think to host-parasite symbiosis. 
Typically it is very difficult to measure the strength of the interactions among two species, so we adopt a standard approach drawing the matrix entries from a given bivariate probability distribution (e.g. Gaussian or Uniform) in the same spirit as traditionally done \cite{may1972,allesina2012stability, grilli2017feasibility}.

We consider a well mixed system, where spatial effects can be neglected. This assumption allows us to obtain some analytical insights on the ecosystem dynamics. 
The dynamics is described by a continuum time stochastic Markov process: a randomly chosen individual is removed and substituted by an individual of the $j$-th species at a rate
\begin{equation}\label{mfr1}
\omega(j,\eta,M,L)=\bar{\eta}^j+\epsilon_1\sum_{k} \bar{\eta}^kM_{kj} \theta(\bar{\eta}^j)+\epsilon_2\sum_{k} \bar{\eta}^kL_{kj}\bar\eta^j
\end{equation}
where $\epsilon_1>0$ and $\epsilon_2>0$ give the cooperation and exploitation intensity, and $\theta(\cdot)$ is the Heaviside step function, i.e., $\theta(x)>0$ when $x>0$ and 0 otherwise. 
The presence of the $\theta$-function in the mutualistic contribution, guarantees that the transition rate is zero if the $j$-th species is extinct. 
For $\epsilon_1=\epsilon_2=0$ we recover the standard VM. 
When $\epsilon_1>0$ the species $j$ is favored by the presence of the other species ($k$ in the summation) to which it is connected and by their population; 
on the other hand $\epsilon_2>0$ allows the possibility that a species exploits (or is exploited by) one or more other species.

It is important to highlight the differences of the contribution on Eq. \ref{mfr1} between exploitative and mutualistic/commensalistic interactions. 
In the first case, the interaction term is quadratic in $\bar{\eta}$ (i.e. $\bar{\eta}^k L_{kj} \bar{\eta}^j$), as exploitative interactions can be derived using the law of mass-action used to describe chemical reactions \cite{matsuda1992statistical, mckane2004stochastic, constable2015models}: a contact must occur between species and the chance of this interaction is, in the simplest hypothesis, proportional to both species concentrations. 
On the other hand, in mutualistic/commensalistic relationships, the contribution to the birth rate is linear in $\bar{\eta}$ (i.e. $\bar{\eta}^kM_{kj}$). 
Indeed, mutualistic interactions (e.g. cross-feeding) are typically mediated by some resource for the species $j$ produced by the species $k$ and proportional to its abundance (e.g. pollen, faecal pellets, metabolic waste) and we assume that these resources are always fully utilized in the community. 
In this setting, what really contributes to the birth rate of a given species is the amount of proper resources in the environment. 
Therefore, the benefit that a species receives does not depend on its own abundance (that is limited by that resource), but only on the abundance of the mutualistic partner. 
To show this we provide a derivation of the linear contribution to the birth rate given by mutualism (see Supporting Information, section 1). While we explicitly consider the case of microbial communities, this could also be extended to other mutualistic systems (e.g. plant-pollinators ecological communities). 

The microscopic dynamics given by rates Eq. \ref{mfr1} induces a Markovian evolution on the relative abundance $\bar{\eta}^s$ of each species. 
Standard techniques \cite{ethier2009markov} can be used to prove that as $N \to \infty$, the process $(\bar{\eta}^1(t),\ldots,\bar{\eta}^S(t))_{t \geq 0}$ weakly converges to the solution of the system of ordinary differential (mean field) equations:
\begin{align}\label{dlmfL}
&\dfrac{\mathrm{d}}{\mathrm{d}t}\bar{\eta}^s(t) = 
\epsilon_1\:\sum_{k=1}^S\bar{\eta}^k(t) M_{ks}\:\theta(\bar\eta^s(t))+\epsilon_2\:\sum_{k=1}^S\bar{\eta}^k(t) L_{ks}\bar \eta ^s(t)\nonumber  \\
& -\bar{\eta}^s(t)\sum_{i,k=1}^S\Big(\epsilon_1\: \bar{\eta}^k(t) M_{ki}\:\theta(\bar\eta^i(t))
+\epsilon_2\: \bar{\eta}^k(t) L_{ki}\: \bar{\eta}^i(t)\Big)
\end{align}
for $s=1,\ldots,S$, where $\sum_{j=1}^S\bar{\eta}^j(t)=1$ and is conserved by the dynamics.

All presented results do not change when the hard constraint of total fixed population size is relaxed by introducing the possibility for a site to become empty: all stationary populations are simply rescaled by a global multiplicative factor, which depends on $\lambda$ (see Supporting Information, section 2). 
We will show below that, under suitable hypothesis, a stationary solution of Eq. \ref{dlmfL} exists and it will be denoted $m_j =\lim_{t \to \infty} \bar{\eta}^j(t)$.

\subsection{Emergent Ecological Patterns}

Through Eq. \ref{dlmfL} we can study many ecosystem properties of interest. 
One of the most important and studied emergent pattern in ecology, which we can determine within our model, is the relative species abundance (RSA) \cite{Hubbell2001, vallade2003analytical, volkov2003neutral, azaele2016statistical}. 
It describes commonness and rarity of species, thus characterizing the biodiversity of an ecological community. 
In our model, the RSA is given by the mean field stationary solution $(m_1,\dots , m_S)$, which in turn depends on the species interaction matrix $M$ and $L$. 

\begin{figure}[tbhp]
	\centering
	\includegraphics[width=0.9\columnwidth]{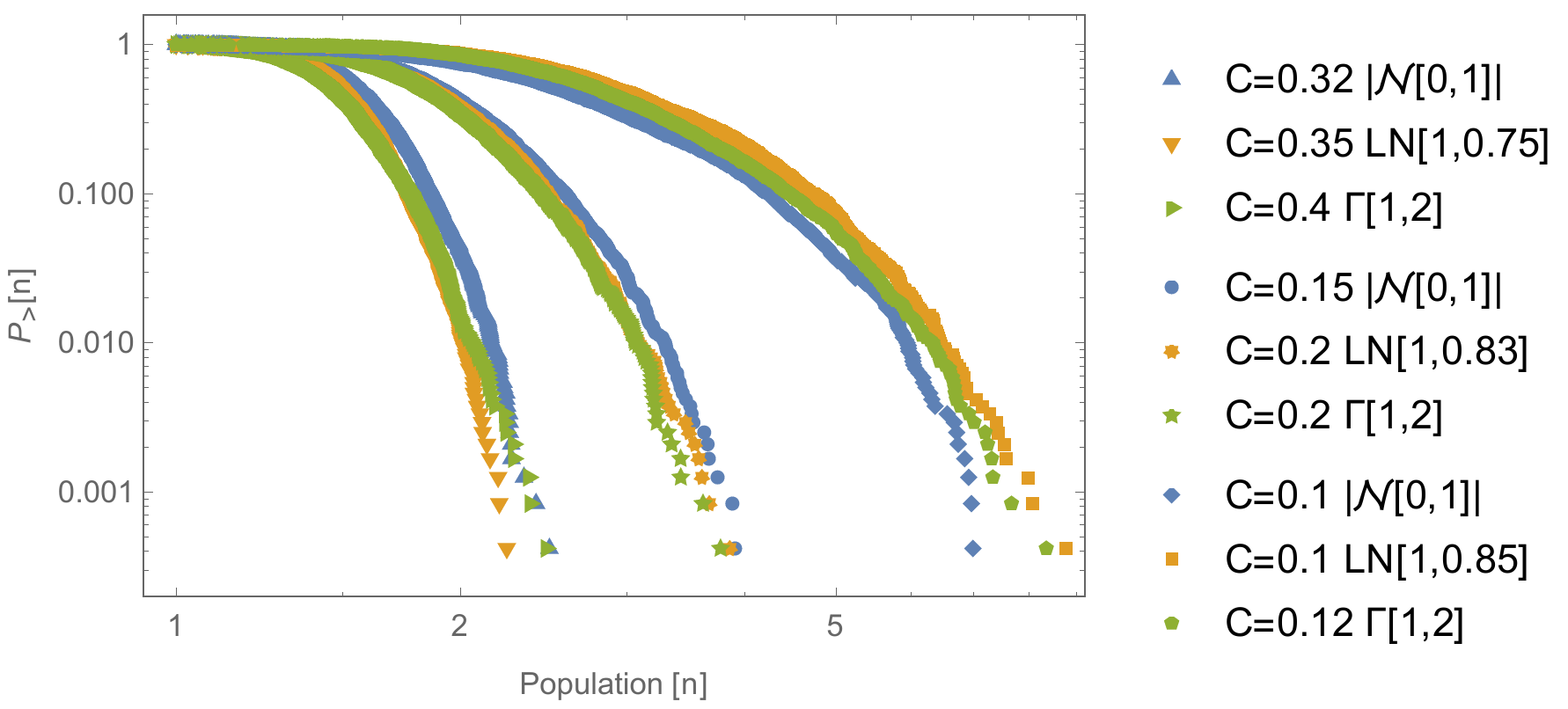}
	\caption{Cumulative RSA for a network of $9$ species, where matrix elements of both $M_{ij}$ and $L_{ij}$ have been drawn from three different probability distributions ($z_h \sim p_h(z)$, $M_{ij}\sim z_h$, $L_{ij}\sim z_h$, $L_{ji}\sim -z_h$, $h=1,2,3$): the modulus of a Normal distribution $z_1\sim|\mathcal{N}(\alpha,\beta)|$ (blue lines), Gamma distribution $z_2\sim\Gamma(\alpha,\beta)$ (green lines) and LogNormal distribution $z_3\sim LN(\alpha,\beta)$ (orange lines lines). Plots display averages over 100 realizations (assuming self-averaging this is equivalent to consider large $S$). Connectivity for mutalistic interaction ($M$) is denoted by $C_M=C$, while for exploitative interactions is $C_L=0.1 C$ (in all the studied cases $C_M+C_L \leq 1$). $\epsilon_1 = \epsilon_2 = 1$. We set the distribution parameters $\alpha,\beta$ (see legend) so that in each case we build interaction matrices with three different values of coefficient of variation $\text{CV} \approx 2,3,4$. As we can see, the cumulative RSA is not very sensible to the distribution from which the matrix elements of both $M_{ij}$ and $L_{ij}$ are drawn, but only on the CV. The analytical formula of $\mu_M, \mu_L, \sigma^2_M,\sigma^2_L$, which depend the network size, connectivity and correlations are presented in the Methods section.}
	\label{FIG1FINAL}
\end{figure}

The cumulative RSA is thus defined as the fraction of species with population greater that a certain value $n$,
\begin{equation}
P_>[n]= \frac{1}{S}\sum_{k=1}^S \theta(n- Nm_k),
\end{equation}
where we have fixed $N=1/\min \{ m_1, \ldots, m_S\}$ when all species coexist, i.e. we have made the choice that the rarest species has population equal to $1$.
We numerically find that the stationary RSA displays a log-normal shape, as the one found in many real ecosystems \cite{azaele2016statistical}, and weakly depends on the specific distribution of the matrix elements $M_{ij}$ and $L_{ij}$, and it is mainly determined only on its coefficient of variation, $\text{CV} =  \sigma_{M+L} / \mu_{M+L} = (\sqrt{\sigma^2_M+\sigma^2_L}) /(\mu_M+\mu_L)$, i.e. the variability of the interaction strengths relative to the mean of $M+L$ (see Fig. \ref{FIG1FINAL}). 
This allows to constrain the model parameters: in order to parametrize species interactions strengths, that are typically unknown \cite{allesina2012stability, suweis2013emergence}, we can make use of a random matrix approach, where we fix the mean and the variance according to the desired RSA one needs to fit. 


We define species abundance fluctuations as $x^i_N(t)=\sqrt{N}\left(\bar\eta^i(t)-m_i\right)$ for $i=1,\ldots,S$.
Another relevant quantity characterizing the ecosystem biodiversity is the covariance matrix $V$, $V_{ij} = \langle x^i(t)x^j(t)\rangle - \langle x^i(t)\rangle\langle x^j(t)\rangle$, describing the correlations in the population abundance fluctuations between pairs of species population abundances \cite{volkov2009inferring}. 
In our setting we can compute analytically this quantity in the limit of normal fluctuations. 
The stochastic process $\left(x^1_N(t),\ldots,x^S_N(t)\right)$ converges in distribution to a Gaussian Markov process $X:=\left(X^1(t),\ldots,X^S(t)\right)$, which solves the stochastic differential equation 
$dX=AX\:dt+\Phi dB_t$, where $B_t$ is a $S$-dimensional Brownian motion, which corresponds to a $S$-dimensional Ornstein-Uhlenbeck process \cite{ethier2009markov, gardiner2004handbook}. 
The analytical expressions for the matrices $A$ and $\Phi$ in terms of the interaction matrices $M$ and $L$, and of the equilibria, $(m_1,\dots, m_S)$, of Eq. \ref{dlmfL}, are given in the Supporting Information, section 3. 
The covariance matrix, $V$, can be obtained by solving the following Lyapunov matrix equation $A\:V+V\:A^T+\Phi \Phi^T=0$.

This quantity is typically measured from species population time series, through the Pearson (or other type of) correlations \cite{faust2012microbial}. 
Moreover, in many studies once opportunely thresholded, it is used as an empirical proxy of the species interactions matrix \cite{faust2012microbial, lima2015determinants}. 
In other words many works assume that $L+M$ can be approximated through $V$. 
Other works, applying maximum entropy approach, use $V^{-1}$ as the quantity to describe the species interactions network \cite{volkov2009inferring}. 
However we find that both $V$ and $V^{-1}$ are not good proxies of the species interactions matrix $M+L$ (see Supporting Information, section 4). 
This result highlights the importance to properly infer interaction networks from data \cite{faust2012microbial, fort2015predicting} by considering a suitable model, which explicitly takes into account species interactions.

\subsection{The importance of cooperation: a solution of the stability-complexity paradox}

We now show how our shift in the assumptions behind mutualistic/commensalistic species interactions could resolve the problematic aspect of stability in ecosystem dynamics. 
In particular, for $\epsilon_1=\epsilon$ and $\epsilon_2=0$ (voter model with cooperation and indirect competition, but no exploitation) we are able to analytically relate key dynamical features of Eq. \ref{dlmfL} to the topology of the interaction matrix $M$ and prove various results of ecological importance. 

First we show that the presence of non-supported species -- the i-th species is non-supported if $\sum_j M_{ji} =0$ --  inhibits coexistence equilibria of the whole ecological community. 
More precisely, if species $i$ is non-supported by other species then at stationarity Eq. \ref{dlmfL} implies that $m_i=0$. 
The extinction of the $i$-th may create new unsupported species that go to zero in the large time limit. 
Such a cascade of extinctions may eventually end only when $\sum_j M_{ji} >0$ for all nodes/species $i$ of the network (see Supporting Information, section 4). 
The elimination of nodes of the interaction network corresponding to all non-supported species will be called pruning in the following. 

Furthermore, we have found sufficient conditions on the topology of the mutualistic interaction matrix $M$ for the existence of stable stationary states of Eq. \ref{dlmfL}.
In fact, if $M$ is irreducible, i.e. if for any node $i$ we can reach any other node $j$ through a path of oriented links $(k,l)$ such that $M_{kl}>0$, then the Perron-Frobenius (PF) theorem holds \cite{seneta2006non} and it exists a unique non-trivial stationary state $(m_1,\dots , m_S)$ with only positive entries. This solution is proportional to the left eigenvector, $v$, of $M$ corresponding to the eigenvalue of $M$ with the largest modulus, which turns out to be non-degenerate, real and positive \cite{seneta2006non},  denoted by $\alpha$ in the following (and that for brevity we will refer to it as PF eigenvalue). In other word, if $M$ satisfies the PF theorem, then $\alpha$ tell us how the stationary species abundances $m$ are distributed. The corresponding right eigenvector will be denoted by $w$, and it gives information on how press perturbations spread throughout the network \cite{suweis2015effect}. 
All components of both $v$ and $w$ are strictly positive and $m_i= v_i/\sum_k v_k $. 
An example of irreducible matrix $M$ occurs when $M_{ij}>0$ implies $M_{ji}>0$ and the network has a single connected component.  Many networks architectures that have been observed in natural ecological communities satisfy this condition (e.g. hierarchical modular structure  in mutualistic networks\cite{bascompte2007plant}).

\begin{figure}[tbhp]
	\centering
	\includegraphics[width=0.8\columnwidth]{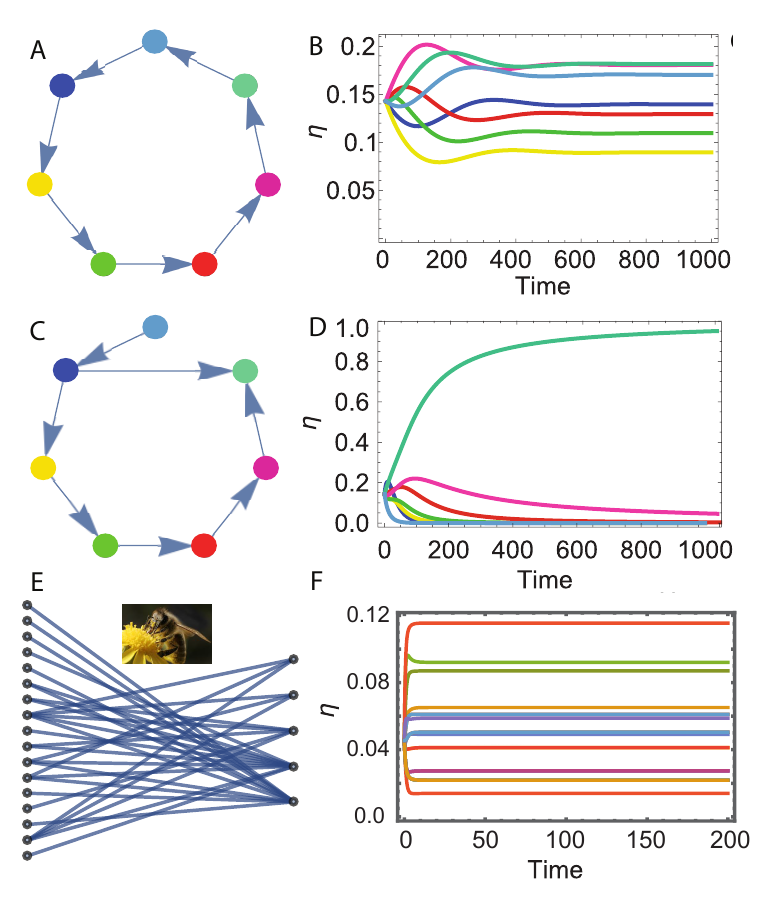}
	\caption{(A) Species interaction network for $7$ species where each species $i$ has one mutualistic partner $j$, i.e. $M_{ij}=1$, $\epsilon_1=1$, $\epsilon_2=0$. (B) Time evolution of the populations of the $7$ species as predicted by the mean field dynamics Eq. \ref{dlmfL}. (C) Species interaction network where one species is not helped by any species and the iterative pruning process, as described in the text, leads to a cascade of extinctions (D) as the time evolution of the mean field Eq. \ref{dlmfL} shows, leading to only one species dominating the community. (E) Nested structure for fruit eating birds community in Mexico \cite{kantak1979observations}. (F) All species coexist, as predicted by our theoretical framework. In the ordinate axis use the notation  $\bar\eta$ and not $\eta$.}
	\label{FIG2FINAL}
\end{figure}

Therefore, within our framework, we can analytically study the impact of the species interaction network architecture on system stability and species extinction. The results of the mean field predictions are shown in Fig. \ref{FIG2FINAL}. Two simple examples are shown corresponding to an ecosystem with no extinction (panels A-B) and with extinction (panels C-D). 

\begin{figure}[tbhp]
	\centering
	\includegraphics[width=0.9\columnwidth]{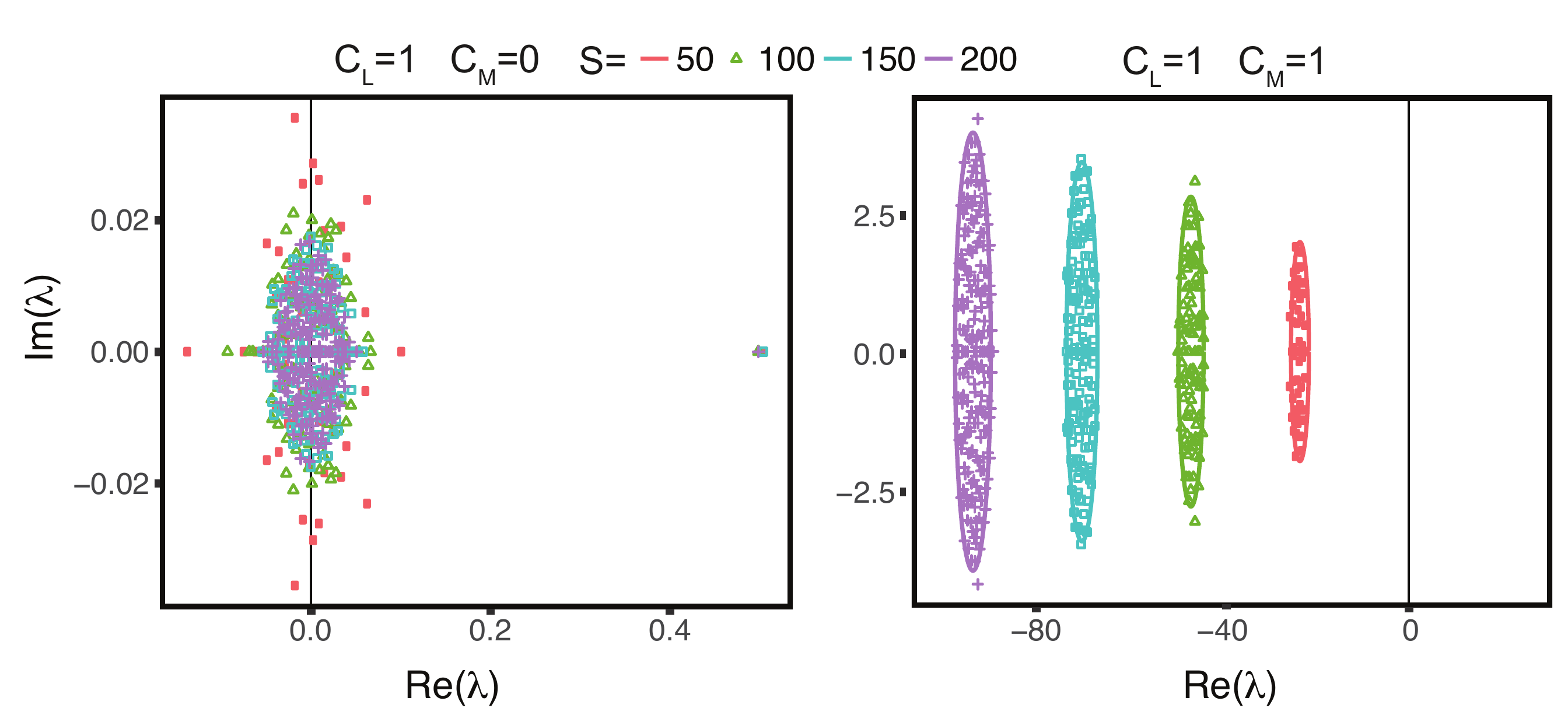}
	\caption{Eigenvalues ($\lambda$) spectrum of the Jacobian matrix $A$ around the stationary state for different size (colors) of the networks (from $S=50$ to $S=200$). The off-diagonal elements of matrices $M$ and $L$ are drawn uniformly between 0 and 1, i.e. $M_{ij}\sim z$, $L_{ij},L_{ji}\sim\pm z$, $z\sim U(0,1)$. Left panel: (A) Pure exploitative interactions ($C_L=1,\epsilon_2=1$,$\epsilon_1=0$); Right panel: (B) Exploitative and mutualistic interactions ($C_L=C_M=0.5,\epsilon_1=\epsilon_2=1$). The points are the eigenvalues of one Jacobian matrix obtained sampling at random the matrices M and L, while the lines indicate the analytical prediction for the support of the $A$ eigenvalues in the corresponding cases (see Eq. \ref{approxJ}). The black vertical line indicates the instability threshold.}
	\label{Fig3FINAL}
\end{figure}

More generally, we can study analytically the stability of the equilibria as a function of ecological complexity, by analyzing the eigenvalues of the linearization of Eq. \ref{dlmfL}, i.e. the Jacobian matrix $A$, around the equilibria, $m_i$, of the system. 
We set equal to zero the diagonal of $M$ whereas the off-diagonal pair $(M_{ij}, M_{ji})$ is equal to $(0,0)$ with probability $1-C$ and with probability $C$ it is drawn from a bivariate Gaussian distribution of means $(\mu,\mu)^T$ and interaction covariance matrix $\Sigma=(\sigma^2 , \rho\sigma^2; \rho\sigma^2 , \sigma^2)$. 
This guarantees that, for a connected cluster, coexistence of all species occurs. 
We define $\mu_M$, $\sigma_M^2$ and $\mu_M=C\mu,\  \sigma_M^2=C\sigma^2+C(1-C)\mu^2,\  \rho_M=\frac{\rho \sigma^2 + (1-C) \mu^2}{\sigma^2 + (1-C) \mu^2}$ as mean, variance and correlation of the elements of matrix $M$.
The case in which each element of $M_{ij}$ is assigned independently of $M_{ji}$ simply correspond to the case $\rho=0$ (notice that even if $\rho=0$ we can have $\rho_M \neq 0$). 
Similarly, when considering also exploitative interactions, we can sample randomly the off-diagonal pairs $(L_{ij}, L_{ji})$, obtaining a given mean $\mu_L$, variance $\sigma_L^2$ and correlation $\rho_L$. 
If $\mu_M \geq \sigma_M \sqrt{(1+\rho_M)/S}$, the leading eigenvalue $\lambda_M = S \mu_M = S C \mu$ and the corresponding eigenvector has positive components \cite{allesina2012stability}. 
Moreover, the components of the leading eigenvector are approximately constant, i.e. the equilibria of system given by Eq. \ref{dlmfL} can be written as $m_i=\frac{1}{S}(1+\xi_i)$ for $i=1,\ldots,S$ with $\sum_i \xi_i = 0$. 
Using the fact that $1 = S m_i - \xi_i$, $\lambda_M = S \mu_M $ and taking into account that all the terms involving $\xi_j$ are sub-leading in $S$, we obtain that the leading term of the system Jacobian does not depend on $L$ (see Methods) and it is equal to:
\begin{equation}\label{approxJ}
A_{ij} = -\delta_{ij} S \mu_M + \left( M_{ij} - \mu_M \right) = -\delta_{ij} S \mu_M + M'_{ij} \ ,
\end{equation}
where $M'_{ij}: = M_{ij}-\mu_M$ is a random matrix with zero mean variance $\sigma_M^2$ and correlation $\rho_M$.
This implies that the eigenvalues are uniformly distributed in an ellipse centered around $-S\mu_M$ with semi-axis $\sqrt{S} \sigma_M (1+\rho_M)$ and $\sqrt{S} \sigma_M (1-\rho_M)$ \cite{girko1985elliptic, sommers1988spectrum}. 
The largest eigenvalue of the Jacobian is therefore given by $-S\mu_M + \sqrt{S} \sigma_M (1+\rho_M)$. 
Thus, for fixed connectivity, $C$, in the presence of cooperation the system stability increases with $S$, whereas if only predator-prey interactions are present, then the stability decreases for increasing ecosystem complexity, as the May theorem would predict (Figure \ref{Fig3FINAL}).

\section{Discussion}

Our results can be applied to study the effect of the interaction network topology to species coexistence in real mutualistic ecological communities. 
In particular, we found that nested architecture \cite{bascompte2003nested}, observed in plant - pollinators ecological communities \cite{bascompte2007plant, suweis2013emergence}, where specialist species, with only few mutualistic links, tend to interact with a proper subset of the many mutualistic partners of any of the generalist species, (see Fig. \ref{FIG2FINAL} panel E) satisfies the hypothesis of the PF theorem and thus favour species coexistence (Fig. \ref{FIG2FINAL} panel F). 

\begin{figure}[h!]
	\centering
	\includegraphics[width=0.9\columnwidth]{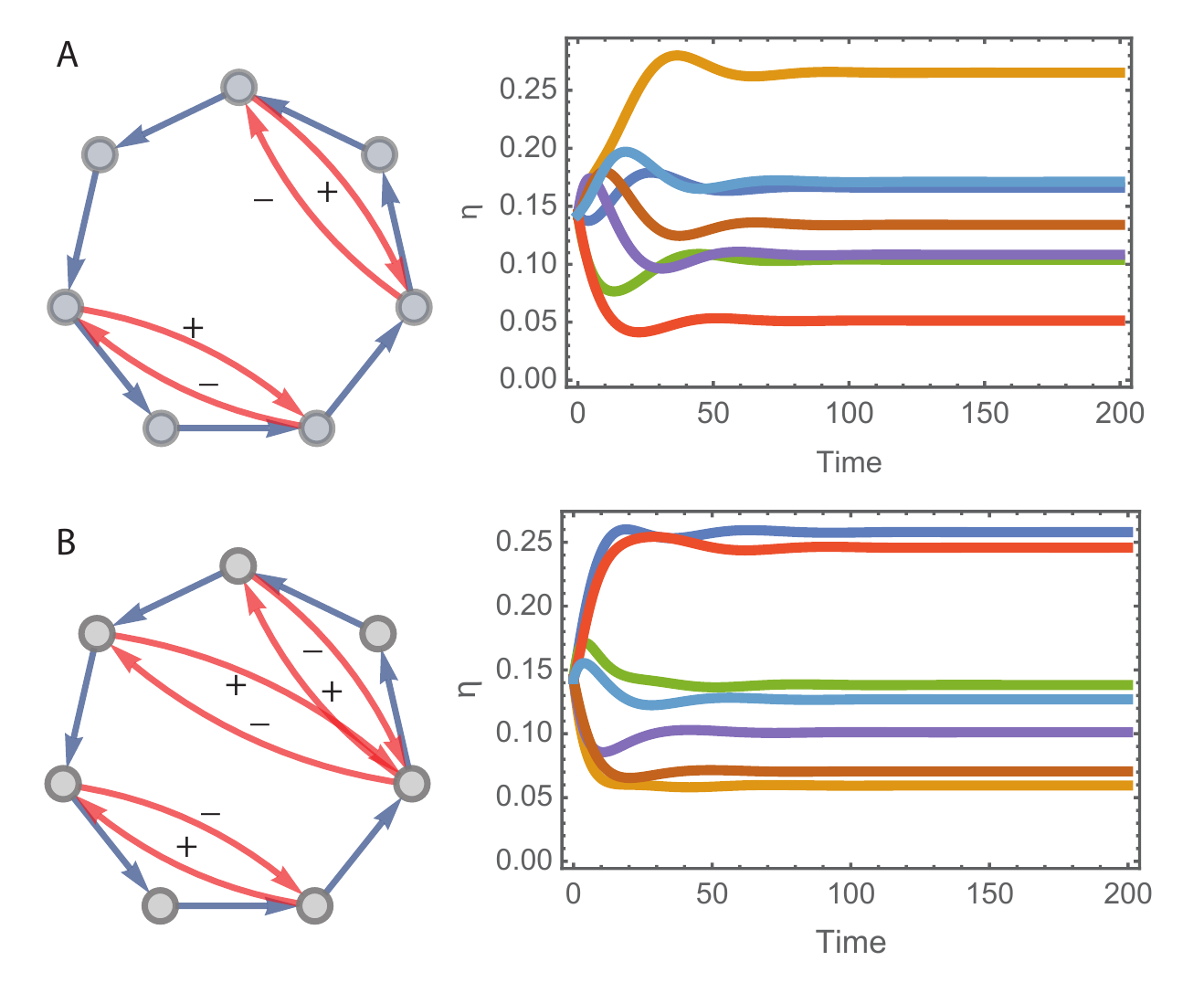}
	\caption{Species interaction network for $7$ species where each species $i$ has one mutualistic partner $j$, i.e. $M_{ij}=1$, and also two (A) and three (B) exploitative (+-) interactions ($\epsilon_1=\epsilon_2=1$). The corresponding time evolution of the populations of the $7$ species, as predicted by the mean field dynamics Eq.  \ref{dlmfL}, are also shown. During the time evolution the rates given by Eq. \ref{mfr1} remain positive and extinctions are not observed.}
	\label{FIGEXPL}
\end{figure}

We have also numerically explored the effect of adding exploitation, i.e. $\epsilon_2\neq 0$ and $C_L>0$ (see Fig. \ref{FIGEXPL}). 
Specifically we find that adding exploitations does not change the main conclusions of our results, as long as a mutualistic network of interactions is present, corresponding to an irreducible matrix, $M$, and the transition rates given by Eq. \ref{mfr1} never become negative during the time evolution of the mean field equation Eq. \ref{dlmfL} (otherwise it would invalidate the derivation of the mean field equations themselves, see Methods section and Supporting Information, section 5).

We have shown that by properly deriving the contribution of mutualism in the species population dynamics, we solve two long standing problems in theoretical ecology: how a large number of species can coexist together and the complexity-stability paradox. 
In fact, we found that cooperation promotes ecosystem biodiversity, that in turn increases its stability without any fine tuning of the species interaction strengths or of the self-interactions \cite{suweis2014disentangling}. 
Even moderate mutualistic interactions can stabilize the dynamics and if present the stability increases with the ecosystem complexity (see Fig. \ref{Fig3FINAL} and Fig. S5-S6 in the Supporting Information). 

Our framework proposes an alternative approach to model cooperation in species population dynamics starting from an individual based stochastic model. 
We have developed a generalization of the classic voter model, adding the effect of species interactions on birth rates. We have shown that a shift in the assumptions behind mutualistic/commensalistic interactions resolve long-standing open theoretical question on the relation between stability and complexity and provides a unifying modeling approach useful to describe emergent patterns in ecology and interacting large ecological systems. 
We highlight that, when properly accounted in the dynamics, mutualistic/commensalistic relationship are crucial in order to have coexistence of species in the communities, as observed recently in real microbial communities \cite{goldford2017emergent,pacheco2018costless}.

\section{Method}

\subsection{Application of the Perron Frobenius Theorem to the Model Equations}

Let us consider the dynamics given by Eq. (\ref{dlmfL}) for $\epsilon_2=0$. If $M$ is irreducible, then the PF theorem holds \cite{seneta2006non} and given the initial condition $\bar{\eta}^i(0) >0\ i=1,\dots,\ S$,  the time dependent solution for the species fractions is
\begin{equation}\label{mfs}
\bar{\eta}(t) = \frac{\bar{\eta}(0)^Te^{\epsilon Mt}}{\sum_i  (\bar{\eta}(0)^Te^{\epsilon Mt})_i}
\end{equation}
Since for any eigenvalue, $\beta \neq \alpha$, of $M$ we have $\Re(\beta) < \alpha$ the dominant term in both numerator and denominator in Eq. \ref{mfs} is $ v \ e^{\alpha t} (\bar\eta(0)\cdot w)$ leading to $\lim_{t \to \infty}\bar\eta(t) = \frac{v}{\sum_i  v_i}= m$. 
This is an easy computation when $M$ has a basis of eigenvectors and in general can be derived using the Jordan decomposition. 
As a corollary of the derivation above we have also that the stationary solution is globally stable in the region $\bar{\eta}^i(0) >0$, for all $i = 1,\ldots,N$.
	
\subsection{Analytical justification of the coexistence condition}
	
As explained in the main text, if the matrix $M$ is irreducible and the transition rates given by Eq. \ref{mfr1} are positive during the time evolution (a necessary condition in order that the derivation of the mean field is justified), then we find numerically that, even in presence of a large concentrations of exploitative interactions, at stationarity the system still admits an high biodiversity and full coexistence is observed (see Fig. \ref{FIGEXPL}). 
Here we want to heuristically justify what we have observed numerically. 
Adding exploitative interactions does not lead to extinctions, as long as the mutualistic network of interactions is present, corresponding to an irreducible matrix, $M$. 
We argue that, under this hypothesis, when $\bar{\eta}^s$ is positive but close to zero the complete mean field equations - where both $\epsilon_1$ and $\epsilon_2$ are positive - are perturbation of the mean field equation where only mutualistic interaction are present,  since we have proved that a pure mutualistic system has no extinction as long as the matrix $M$ is irreducible. 
Following the notation in Results, our continuous time Markov process is defined by the rule: a randomly chosen individual is removed and substituted by an individual of the $j$-th species at a rate
\begin{equation}\label{rates}
\omega(j,\eta,M,L) = \underbrace{\bar{\eta}^j+\epsilon_1\sum_{k} \bar{\eta}^kM_{kj} \theta(\bar{\eta}^j)}_{:=\omega_j^M}+\underbrace{\epsilon_2\sum_{k} \bar{\eta}^kL_{kj}\bar\eta^j}_{:=\omega_j^L},
\end{equation}
where $\epsilon_1>0$ and $\epsilon_2>0$ give the cooperation and exploitation intensity, and  $\theta(\cdot)$ is the Heaviside step function, i.e., $\theta(x)>0$ when $x>0$ and 0 otherwise. 
As $N \to \infty$ the relative abundance $\bar{\eta}^s$ converges to the solution of the system of ordinary differential equation for $s=1,\ldots,S$. 
Equation for $\bar{\eta}^s$, when $\bar{\eta}^s$ is positive but close to zero, can be written in the following form
\begin{equation}\label{equation2}
\dfrac{\mathrm{d}}{\mathrm{d}t}\bar{\eta}^s(t) =\underbrace{\omega^M_s-\bar\eta^s(t)\sum_{i}\omega^M_i}_{\simeq\delta>0}+\mathcal{O}(\bar{\eta}^s) 
\end{equation}
The first two terms in Eq. \ref{equation2} are the vector fields corresponding to mean field equation for $M$ irreducible and no exploitation (i.e. $\epsilon_2=0$). 
We know that such a system has no extinctions and its vector field is typically greater than $\delta>0$ out of equilibrium when $\bar{\eta}^s \simeq 0$. 
The last term in Eq. \ref{equation2} contains terms which are linear dependent of $\omega_j^L$ which is $\mathcal{O}(\bar\eta_s)$.  
Thus $\dfrac{\mathrm{d}}{\mathrm{d}t}\bar{\eta}^s(t)$ is positive for $\bar\eta^s$ close to zero. 
The requested transition rates never become negative during the time evolution of the mean field equation. 
This is a necessary condition otherwise the derivation itself of the mean filed equation would be meaningless.

\subsection{Stability of the equilibria}\label{stabL}
	
In the case of $\epsilon_2 \neq 0$, the entries of the Jacobian read
\begin{align*}
\nonumber A_{ij}&=\epsilon_1\left(M^T_{ij}-\delta_{ij}\sum_{h,k=1}^Sm_hM_{hk}-m_i\sum_{k=1}^SM_{jk}\right) &  \\
\nonumber &+\epsilon_2\left(L_{ji}m_i+\delta_{ij}\sum_{k=1}^Sm_hL_{hi}-\delta_{ij}\sum_{h,k=1}^Sm_hL_{hk}m_k\right.\\
&\left.-m_i\sum_{k=1}^SL_{jk}m_k-m_i\sum_{k=1}^SL_{kj}m_k\right) 
\end{align*}
The diagonal entries of the Jacobian are
\begin{equation}\label{Jac}
A_{ii} = -\epsilon_1 \sum_{h,k=1}^S m_h M_{hk} + \epsilon_2 \sum_{h=1}^S m_h L_{hi} - \epsilon_2 \sum_{h,k=1}^S m_hL_{hk} m_k 
\end{equation}
Since $m_i \sim 1/S$, it is simple to observe that the term proportional to $\epsilon_1$ is of order $S$ (plus sub-leading fluctuations). 
On the other hand, the leading order of the terms proportional to $\epsilon_2$, is of order $1$ and therefore always sub-leading if $\epsilon_1 > 0$. 
A similar argument applies to the off-diagonal elements. 
In that case, the terms proportional to $\epsilon_1$ are of order $1$, while the ones proportional to $\epsilon_2$ are of order $1/S$.
	
Similarly to what found in the case $\epsilon_2 = 0$, we have that the following relations hold: $\mu_{L} = C_L \mu, \; \sigma_{L} = \sqrt{C_L \left( \sigma^2 + (1-C_L) \mu^2 \right)}, \; \rho_{L} = \frac{\rho \sigma^2 + (1-C_L) \mu^2}{\sigma^2 + (1-C_L) \mu^2}$ 
where $\mu$ and $\sigma$ are the mean and the standard deviation of the distribution from which we draw the value for the exploitative interaction strengths. 
These expressions have been used together with $\mu_M$, $\sigma_M$ and $\rho_M$, when calculating the coefficient of variation. 
The above considerations indicate that the distribution of the eigenvalues of the Jacobian, Eq. \ref{Jac}, is the same as the $\epsilon_2=0$ case. 
Figures visualising these results are presented in the Supporting Information (section 6).

\section*{Supporting Information}

\subsection*{Mechanistic interpretation of linear growth rates}

The mutualistic dynamics introduced in the main text assumes that the benefit that a species
receives from other species is independent of its own abundance. This assumption is radically
different from the typical form of growth rates for exploitative (e.g. predator-prey) interactions, where some sort of mass-action law, typical of chemical reaction, is usually invoked \cite{staniczenko2013ghost}.
Here we consider mutualism/commensalism as the presence of certain species is able to create an environment (e.g. by producing some public good or nutrient) or to release some substances (e.g. faecal pellets, metabolic waste), which favor the growth of some others species. 

The state at time $t$ of the system is given by species concentration  vector (i.e., average fraction of individuals for each species) $\bar{\eta}^i(t)$ with $i=1,\ldots,S$ and let $c$ be the concentration of a given resource used by the species $j$. 
This resource is provided, at a rate $s$, by certain species, $k$'s (e.g. through metabolic waste/secretion \cite{goldford2017emergent, taillefumier2017microbial,pacheco2018costless} or, in the case of plant/flowers, it represents the pollen produced by the $k$'s species) and related to their populations in a linear way, that is $s(t)= \sum_k \bar{\eta}^k(t) M_{kj}$. 
The kinetic of nutrient concentration is then \cite{posfai2017metabolic, taillefumier2017microbial}
\begin{equation}
\frac{dc(t)}{dt} = \frac{1}{\tau_R} \left( s(t)- \bar{\eta}^j(t) r(c(t)) \right)
\end{equation}
where $r(c)$ is the consumption rate per individual whose specific form is irrelevant for the purpose of this example
(e.g., one can consider the Monod function $r(c)= \alpha c/(K+c)$, with $\alpha$ and $K$ some suitable constants). The constant $\tau_R$ is the timescale of the dynamics of resources.

The contribution to the growth rate $\Delta \omega_j$ of the $j$-th species, 
due to this nutrient, is 
\begin{equation}
\Delta \omega_j= \epsilon r(c) \bar{\eta}^j 
\end{equation}
where $\epsilon$ is a conversion factor measuring how the nutrient contributes to the biomass of the $j$-th species.
If the nutrient concentration is in quasi-steady state \cite{posfai2017metabolic}, that is $dc(t)/dt=0$, which occurs if it relaxes much faster than populations (i.e. for small $\tau_R$), then, from the above equation, we get
\begin{equation}
r(c(t))= \frac{s(t)}{\bar{\eta}^j(t)}
\end{equation}
leading to 
\begin{equation}
\Delta \omega_j= \epsilon s = \epsilon \sum_k \bar{\eta}^k M_{kj}
\end{equation}
If $\bar{\eta}^j > 0$, then we have that the rate is linear in the population of the mutualistic partners. 
We again highlight that this form is radically different from the typical growth rates proposed in the literature, where some sort of mass-action law, typical of chemical reaction, is usually invoked \cite{staniczenko2013ghost}. 
The latter assumption is in fact only appropriate when we are assuming that the contribution to the population growth depend on the physical encounter between two species, as typically happens for exploitative (e.g. predator-prey) interactions.

\subsection*{Mean field analysis for the voter model with empty sites}

If we turn off exploitation ($\epsilon_2=0$), the mean field equation without empty site ($\epsilon_1=\epsilon$) reads 
\begin{equation}\label{dlmf}
\dfrac{\mathrm{d}}{\mathrm{d}t}\bar{\eta}^s = \epsilon\:\sum_{k=1}^S\bar{\eta}^k M_{ks}\:\theta(\bar\eta^s)-\epsilon\: \bar{\eta}^s\sum_{i,j=1}^S \bar{\eta}^i M_{ij}\:\theta(\bar\eta^j) 
\end{equation}
where $s=1,\ldots,S$ represents different species, $\bar{\eta}^s$ is the average fraction of individuals of the $s$-th species, $M$ is the interaction matrix whose non-zero entries define the network of ecological interactions, $\theta$ is the Heaviside step function ($\theta(x) = 1 (0)$ for $x>0 (x\leq 0)$ ) and $\epsilon$ is the cooperation intensity (the average of the non-zero $M_{ij}$ is fixed to $1$). 
For simplicity, we have omitted time dependence of $\bar{\eta}$.
An intuitive derivation is as follows. 
The key point is that for $N$ large the evolution of the quantity $\bar{\eta}^s$ becomes deterministic because the noise is canceled in the macroscopic regime and in the thermodynamics limit the relative abundance converges to its mean. 
Then, observe that the dynamics of the relative abundance in the infinitesimal time $dt$ is simple as it can only decrease by $1/N$ when a site of kind $s$ change type or can increase by $1/N$ when the new symbol of a certain site is $s$.

We now extend the model presented in the main text introducing the possibility for a site to be empty. 
In our setting empty sites do not interact with species. 
Thus the species rates remain unchanged after the introduction of empty sites. 
Thus the species rates are the same as before whereas non-empty sites become empty with rate $\lambda$. In the case $\epsilon=0$, the rate $\lambda$ has to be less than 1 otherwise empty sites will cover all the available space. 
The mean field equations become now:

\begin{align}\label{dlmfES}
\dfrac{\mathrm{d}}{\mathrm{d}t}\bar{\eta}^s &= \bar{\eta}^s \bar{\eta}^0 - \bar{\eta}^s \lambda + \epsilon \sum_{k=1}^S \bar{\eta}^k M_{ks} \theta(\bar{\eta}^s) - \epsilon \bar{\eta}^s \sum_{i,j=1}^S \bar{\eta}^i M_{ij} \theta(\bar\eta^j) \\
\dfrac{\mathrm{d}}{\mathrm{d}t}\bar{\eta}^0 &= (1-\bar{\eta}^0)(\lambda-\bar{\eta}^0) - \epsilon \bar{\eta}^0 \sum_{i,j=1}^S \bar{\eta}^i M_{ij} \theta(\bar{\eta}^j)
\end{align}

Let us analyze the stationary mean-field equations for $\epsilon << 1$. 
In this case the stable equilibrium for the empty sites is $\bar{\eta}^0 = \lambda - \epsilon \: \frac{\lambda}{1-\lambda} \sum_{i,j=1}^S \bar{\eta}^i M_{ij} \: \theta(\bar{\eta}^j) + O(\epsilon^2)$. Substituting in the equations for $\bar{\eta}^s$, we obtain
\begin{equation}\label{dlmfes}
\epsilon \sum_{k=1}^S \bar{\eta}^k M_{ks} \theta(\bar{\eta}^s) - \epsilon \left(\frac{1}{1-\lambda}\right)  \bar{\eta}^s \sum_{i,j=1}^S \bar{\eta}^i M_{ij} \theta(\bar{\eta}^j) + O(\epsilon^2) = 0
\end{equation}
where $s=1,\ldots,S$.
After the change of variable $\bar{\eta}' = (1-\lambda) \bar{\eta}$, the above \eqref{dlmfes} reduces to the same equation as one would get for $\lambda=0$, i.e. in absence of empty sites the mean field equation becomes:

\begin{align}\label{dlmfL}
\dfrac{\mathrm{d}}{\mathrm{d}t}\bar{\eta}^s(t) = \epsilon_1 \sum_{k=1}^S \bar{\eta}^k(t) M_{ks} \theta(\bar\eta^s(t)) +\epsilon_2 \sum_{k=1}^S \bar{\eta}^k(t) L_{ks} \bar{\eta}^s(t) - \bar{\eta}^s(t) \sum_{i,k=1}^S \Big(\epsilon_1 \bar{\eta}^k(t) M_{ki} \theta(\bar{\eta}^i(t)) + \epsilon_2 \bar{\eta}^k(t) L_{ki} \bar{\eta}^i(t) \Big)
\end{align}

for $s=1,\ldots,S$, where $\sum_{j=1}^S\bar{\eta}^j(t) = 1$ and is conserved by the dynamics. 

In other words, when $\epsilon$ is small, the introduction of empty sites leads to stationary abundances which are trivially rescaled with respect to the case in absence of empty sites, as a consequence of the reduction of the available space.

\subsection*{Covariance matrix and Species Interaction Networks}

In this section, we consider the normal fluctuations around the deterministic limit of Eq. \ref{dlmfL}. 
This allows us to calculate the matrix $V$ describing the correlation between pairs of species population abundances \cite{volkov2009inferring}. 
As highlighted in the main text, this quantity, once opportunely thresholded, is used as an empirical proxy of the species interactions network \cite{faust2012microbial, friedman2012inferring, lima2015determinants}. 
Other works, applying maximum entropy approach, use $V^{-1}$ as the quantity to describe species interactions \cite{volkov2009inferring, stein2015inferring}. 
The aim of this section is to test how well $V$ or $V^{-1}$ approximate the true interactions described by $M+L$ in our model.

For sake of simplicity, we assume that the limiting dynamics start at the equilibrium $m_1,\ldots,m_S$ with $0< m_i <1$, $i=1,\ldots,S$. 
Thus, we define the fluctuation process as
\begin{equation}
x^i_N(t)=\sqrt{N}\left(\bar\eta^i_N(t)-m_i\right) \mbox{ for } i=1,\ldots,S.
\end{equation}
One can apply standard techniques of convergence of generators to get weak convergence to the thermodynamic limiting evolution \cite{ethier2009markov}.
Indeed, the stochastic process $\left(x^1_N(t),\ldots,x^S_N(t)\right)$ converges in distribution to a Gaussian Markov process $X:=\left(X^1(t),\ldots,X^S(t)\right)$ which solves the stochastic differential equation
\begin{equation}\label{flu}
dX=\:AX\:dt+\Phi dB_t
\end{equation}
where $B_t$ is a $S$-dimensional Brownian motion and 

\begin{align*}\label{Acompete}
A_{ij} & = \epsilon_1 \left(M^T_{ij}-\delta_{ij} \sum_{h,k=1}^S m_h M_{hk} - m_i \sum_{k=1}^S M_{jk} \right) \\
&+ \epsilon_2 \left(L_{ji} m_i+\delta_{ij} \sum_{h=1}^S m_hL_{hi}- \delta_{ij} \sum_{h,k=1}^S m_hL_{hk} m_k-m_i \sum_{k=1}^S L_{jk} m_k - m_i\sum_{k=1}^SL_{kj}m_k\right); 
\end{align*}
\begin{align*}
(\Phi\Phi^T)_{ij} &= -2 \left(m_i m_j (1+\epsilon_1 m_i \sum_{h,k=1}^S m_k M_{kh} + \epsilon_2 \sum_{h,k=1}^S m_k L_{kh} m_h)\right) (1-\delta_{ij})  \\
&+ 2 (1-m_i) \left(m_i+\epsilon_1 \sum_{k=1}^S m_k M_{ki} + \epsilon_2 \sum_{k=1}^S m_k L_{ki} m_i \right) \delta_{ij}  
\end{align*}\label{Phicompete}
where $i,\ j=1,\ldots,S$ and $\delta_{ij}$ is the Kronecker delta. 

From Eq. \ref{flu}, it is then possible to derive the dynamics of the covariance matrix (see \cite{gardiner1985stochastic} for details):

\begin{equation}
V_{ij}(t) = \langle X^i(t)X^j(t)\rangle - \langle X^i(t)\rangle \langle X^j(t)\rangle
\end{equation} 
Therefore, we have
\begin{equation}\label{cor1}
\frac{d\: V(t)}{dt}= \:A\:V(t) + \:V(t)\:A^T + \Phi\Phi^T,
\end{equation}
and at stationarity the covariance matrix, $V_{ij} = \lim_{t \to \infty} V_{ij}(t)$, resolves the following equation
\begin{equation}\label{cor2}
\:A\:V + \:V\:A^T + \Phi\Phi^T = 0.
\end{equation}
Eq. \ref{cor2} is a Lyapunov equation, so we could apply standard algorithms to solve it numerically \cite{penzl1998numerical}.

\begin{figure}[h!]
	\centering
	\subfigure[]{
		\includegraphics[width=0.48\columnwidth]{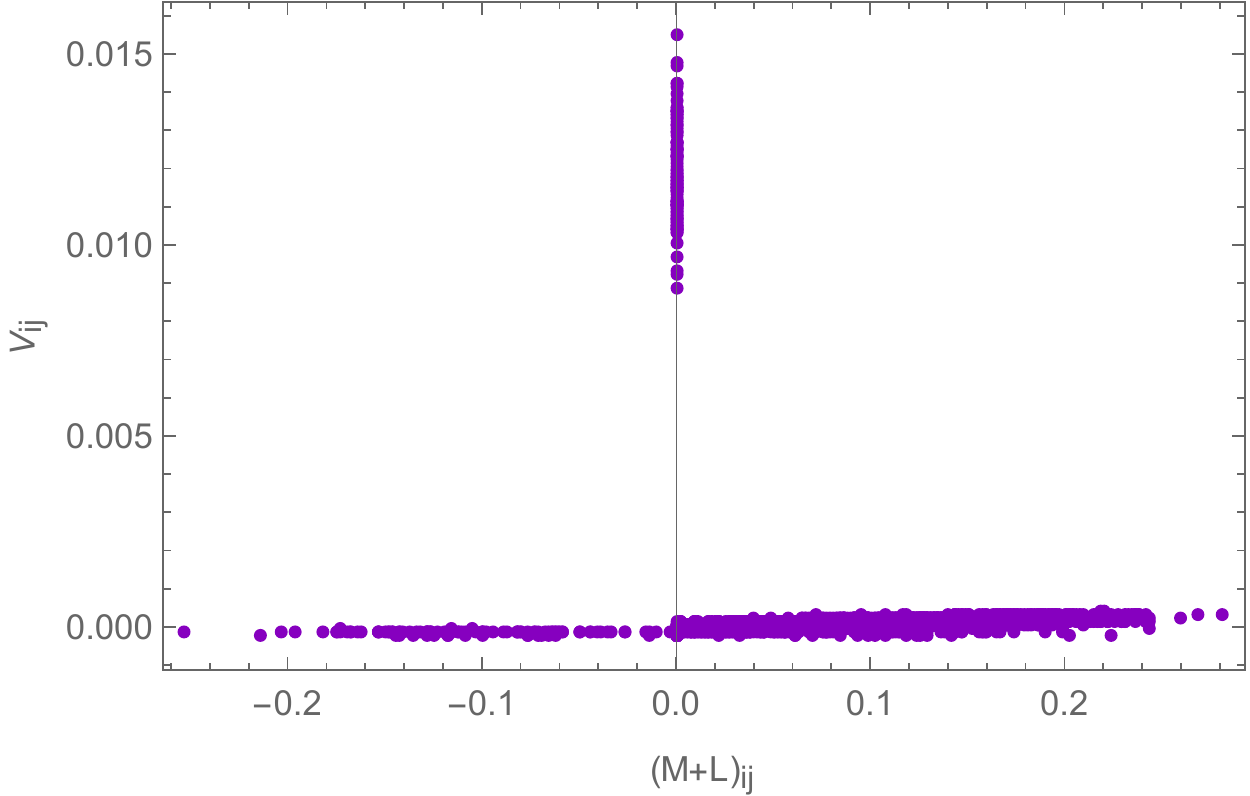}}	
	\subfigure[]{
		\includegraphics[width=0.48\columnwidth]{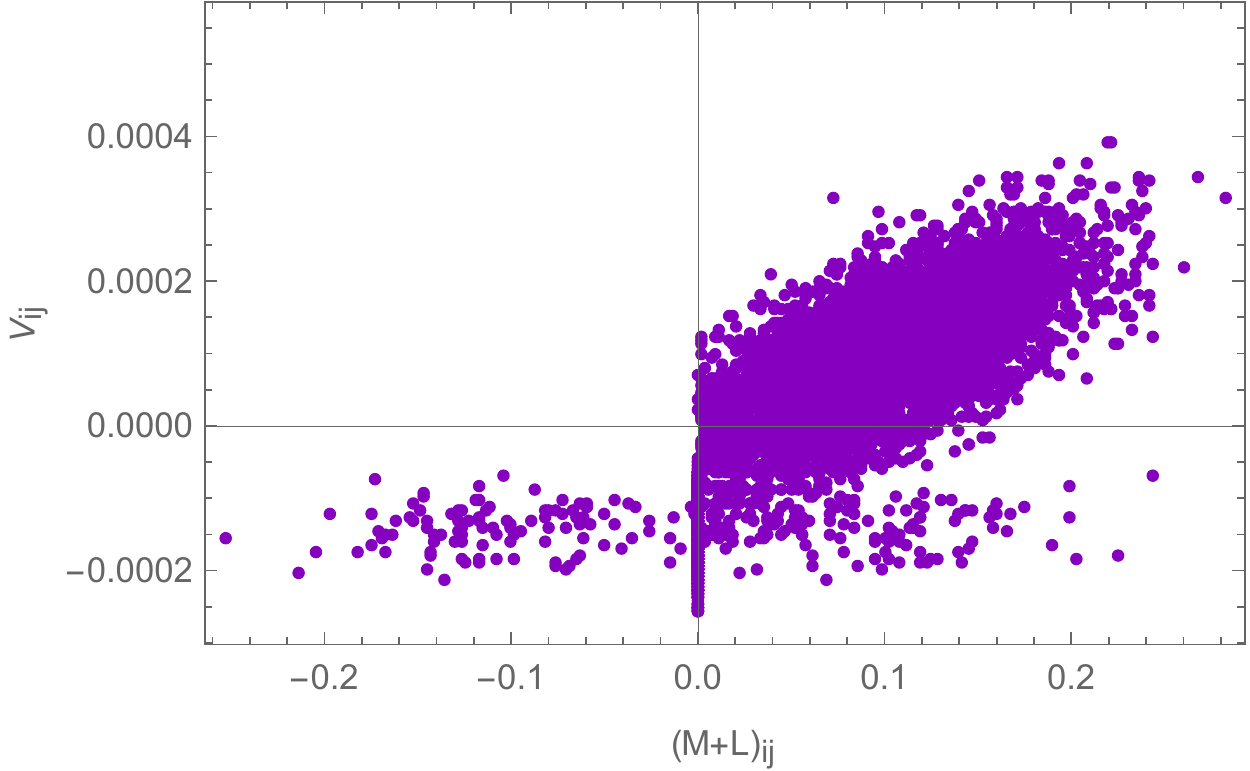}}
	
	\subfigure[]{
		\includegraphics[width=0.48\columnwidth]{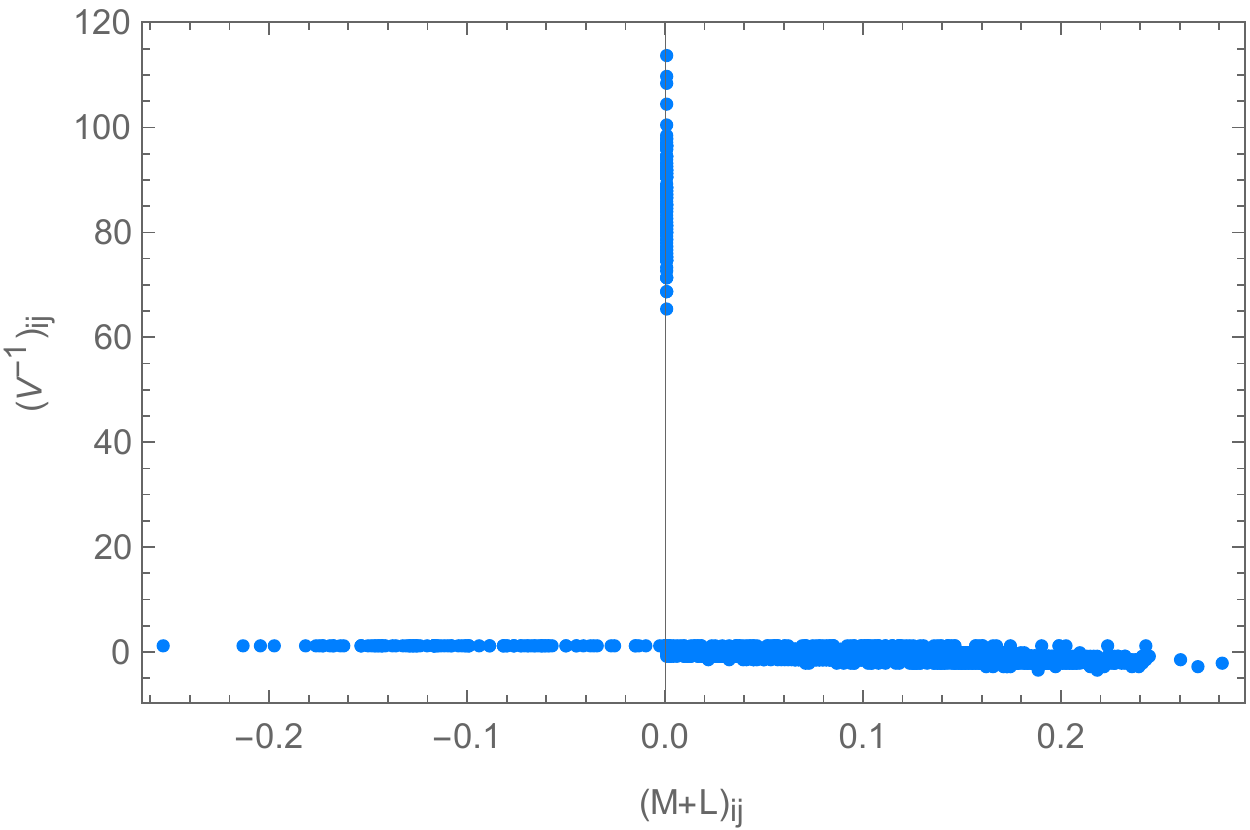}}	
	\subfigure[]{
		\includegraphics[width=0.48\columnwidth]{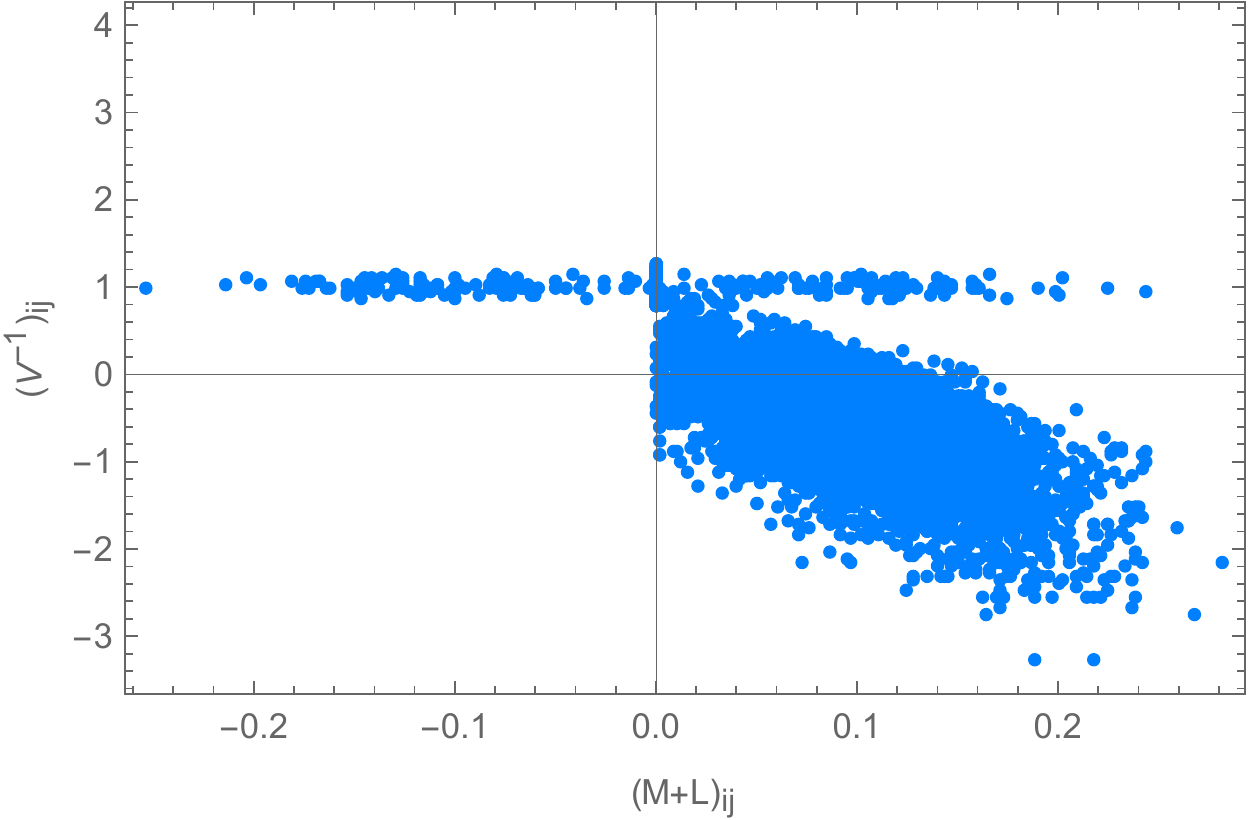}}
	
	\caption{Elements of the covariance matrix $V$ and its inverse $V^{-1}$ compared to the species interaction network $M+L$ with size $S=100$, for dense mutualism $C_M=0.5$ and sparse exploitation $C_L=0.05$. Interaction strengths $z_{ij}$ have been drawn from a Gaussian distribution of mean $\mu_N=0.1$ and standard deviation $\sigma_N=0.05$. The sign has been then chosen accordingly ($M_{ij}=|z_{ij}|$; $L_{ij} = |z_{ij}|$ and $L_{ji} = - |z_{ji}|$). We have also imposed the irreducibility of $M$. Panels $(a),(c)$ represent the correlation over the whole of $L+M$ (between -0.3 and +0.3), while panels $(b),(d)$ zoom in the the relation close to the intersection of the $x$-$y$ axes. Although the zoom highlight a slightly positive (panel $b$) and negative (panel $d$) correlation between elements of $M+L$ and $V$, $V^{-1}$, they are not significant. Most of the elements of both the covariance matrix $V$ and its inverse $V^{-1}$ are close to zero. Other elements are very large, although the corresponding species do not interact (L+M=0), indicating that $V$ or $V^{-1}$ cannot be used as interaction matrix.}\label{fig:CoPPWV}
\end{figure}

We have determined $V$ from the solution of Eq. \ref{cor2} and determined $V^{-1}$. 
If one assume that the population fluctuations around their means are gaussian distributed, then $V^{-1}$ represents the species interaction matrix \cite{lezon2006using, volkov2009inferring}. 
Indeed, within a maximum entropy approach, $V^{-1}$ is typically used to infer species interactions based on the available information of the system \cite{stein2015inferring}. 
In our framework and as shown by Eq. \ref{cor1} and Eq. \ref{cor2}, the relation between the interaction matrix $M+L$ and the matrix $V$ or $V^{-1}$ is highly non-linear. 
Moreover, because of the constraint, $\sum_{j}V_{ij}=0$, $V$ is not invertible, and thus in order to compute $V^{-1}$ we apply a pseudo-inverse scheme, i.e. we invert $V$ is the subspace of spanned by the eigenvectors corresponding to non-zero eigenvalues. 
As shown in Fig. \ref{fig:CoPPWV}, even for very simple structure of matrix $M$ and $L$, $V$ and $V^{-1}$ are not good proxies of the species interactions. 
The results are shown for the model without empty sites, but there is no qualitatively difference with the model including empty sites. 
This result highlights the importance to properly infer interaction networks from data. 

\subsection*{Topology of the Interaction Networks, Coexistence and Stationary States}

In this section, we discuss some features of the topology of the mutualistic interaction matrix $M$ and how they relate to stationary states of the system. The main concept in this section is the one of \emph{pruned} graph and the operation of \emph{pruning} a network. 
A node with in-degree equals to zero and out-degree different from zero is called a \emph{dead leaf} of the network. 
The operation of pruning consists in eliminating one by one the dead leaves of a given network together with their outbound links. 
After a first pruning, we will obtain a new network (that is a subnetwork of the starting one) that may still have dead leaves - the elimination of dead leaves may create new dead leaves. 
The pruning process end when the resulting network has no more dead leaves.
The latter network is called \emph{stable} or \emph{pruned}. 
It is easy to see that the minimal pruned network (i.e. with the smallest number of links) that can be constructed with $S$ nodes is the cyclic graph. 
More in general, we have:\\

\noindent{\textbf{Proposition:}} The pruned network is a union of isolated nodes and graphs that contain at least one cycle each.\\

Indeed, pruning stops when the obtained graph is a union of isolated nodes and graphs where all nodes have at least an ancestor (i.e. the in-degree of each node is positive). 
Now a finite graph where each node has a least one incoming link contains at least a cycle. 
In fact, starting from one node it is possible to walk through the ancestors and never stop. 
Since the graph is finite, soon or later, the walker will visit twice the same node - so the walk contains a cycle - at most after a number of steps that equals the size of the graph.

\begin{figure}[h!]
	\centering
	\scalebox{0.7}{
		\begin{tabular}{cc}
			\tikzset{every node/.style={%
					draw, circle, minimum size=5mm},
				help lines/.append style=pink}
			\begin{tikzpicture}
			\coordinate [label=left:$(a)$] (a) at (-2.5,2.5);  
			\draw[blue] (a); 
			\node (0) at (-2.5,0.5) {$0$};
			\node (1) at (-0.5,0.5) {$1$};
			\node (2) at (0.5,2.5) {$2$};
			\node (3) at (1.5,0.5) {$3$};
			\node (4) at (2.5,2.5) {$4$};
			\draw [-latex,red,thick, bend right] (2) to (3);
			\draw [-latex,red,thick, bend right] (3) to (4);
			\draw [-latex,red,thick, bend right] (4) to (2);
			\draw [-latex] (1)  to (3);
			\draw [-latex] (0)  to (1);
			\coordinate [label={[black]center: $\times$}] (C) at (0.5,0.5);
			\coordinate [label={[black]center: $\times$}] (C) at (-1.5,0.5);
			\end{tikzpicture}&
			
			\tikzset{every node/.style={%
					draw, circle, minimum size=5mm},
				help lines/.append style=pink}
			\begin{tikzpicture}
			\coordinate [label=left:$(b)$] (b) at (-1.9,2.5);  
			\draw[blue] (b);
			\node (1) at (-0.5,0.5) {$1$};
			\node (2) at (0.5,2.5) {$2$};
			\node (3) at (1.5,0.5) {$3$};
			\node (4) at (2.5,2.5) {$4$};
			\node (5) at (1.5,1.7) {$5$};
			\node (6) at (-.5,3.5) {$6$};
			\node (7) at (-.5,2.5) {$7$};
			\node (8) at (-1.5,2.5) {$8$};
			\draw [-latex,red,thick, bend right] (2) to (3);
			\draw [-latex,red,thick, bend right] (3) to (4);
			\draw [-latex,red,thick, bend right] (4) to (2);
			\draw [-latex,red,thick,bend right] (5) to (4);
			\draw [-latex,red,thick,bend right] (4) to (5);
			\draw [-latex,red,thick] (5) to (2);
			\draw [-latex,red,thick] (5) to (3);
			\draw [-latex,red,thick] (5) to (3);
			\draw [-latex,red,thick] (5) to (3);
			\draw [-latex,red,thick] (3)  to (1);
			\draw [-latex,red,thick] (2)  to (6);
			\draw [-latex,red,thick] (6)  to (8);
			\draw [-latex,red,thick] (1)  to (7);
			\draw [-latex,red,thick] (7)  to (2);
			\end{tikzpicture}\\
		\end{tabular}
	}
	\caption{The diagram of how to prune network. (a) An example of how the operation of pruning works. First the 0-node is eliminated with its outbound link. After that, the node 1 becomes a dead leaf and has to be pruned. The cycle shown by the red links is the resulting pruned network. (b) An example of a pruned network that is not composed only by cycles.}\label{fig:PruningNet}
\end{figure}
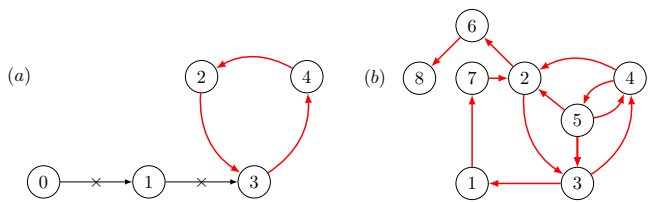

The pruned network has at least one cycle but when not simply union of isolated cycles it can be very complex. 
Fig. \ref{fig:PruningNet} shows an example of the pruning procedure and of a non-trivial pruned network.	  

As we anticipated at the beginning of this section, the dynamics of species sitting on dead leaves of the interaction network is trivial as their relative abundance goes to zero. 
This is a simple consequence of the fact that a dead leaf has no incoming bond. 
Thus, when $s$ is a dead leaf, the first term on the right of Eq. \ref{dlmf} is zero and simple estimate gives $\mathrm{d}\bar{\eta}^s/\mathrm{d}t = -\epsilon \:\bar\eta^s\sum_{i,j}\bar\eta^{i}M_{i j}\theta(\bar\eta^j) \leq 0$.
The previous simple remark leads to the following:\\

\noindent{\textbf{Limiting dynamics of dead leaves:}} Start the dynamics from a point with $\bar\eta^i\neq 0$ for all $i=1,\ldots,S$. If $k$ is a dead leaf then $\lim_{t \to \infty} \bar{\eta}^k(t) = 0$.\\

Thus the presence of a dead leaf inhibits coexistence equilibria on the whole graph. 
More precisely,  if $i=1,\ldots,\gamma$ are dead leaves (at some step of the pruning), the stable equilibria must have $\bar{\eta}^1 = \ldots = \bar{\eta}^{\gamma} = 0$.

\subsection*{Mean Field Equations, Birth Rates and Species Coexistence}

\begin{figure}[h!]
	\centering
	\includegraphics[width=0.85\columnwidth]{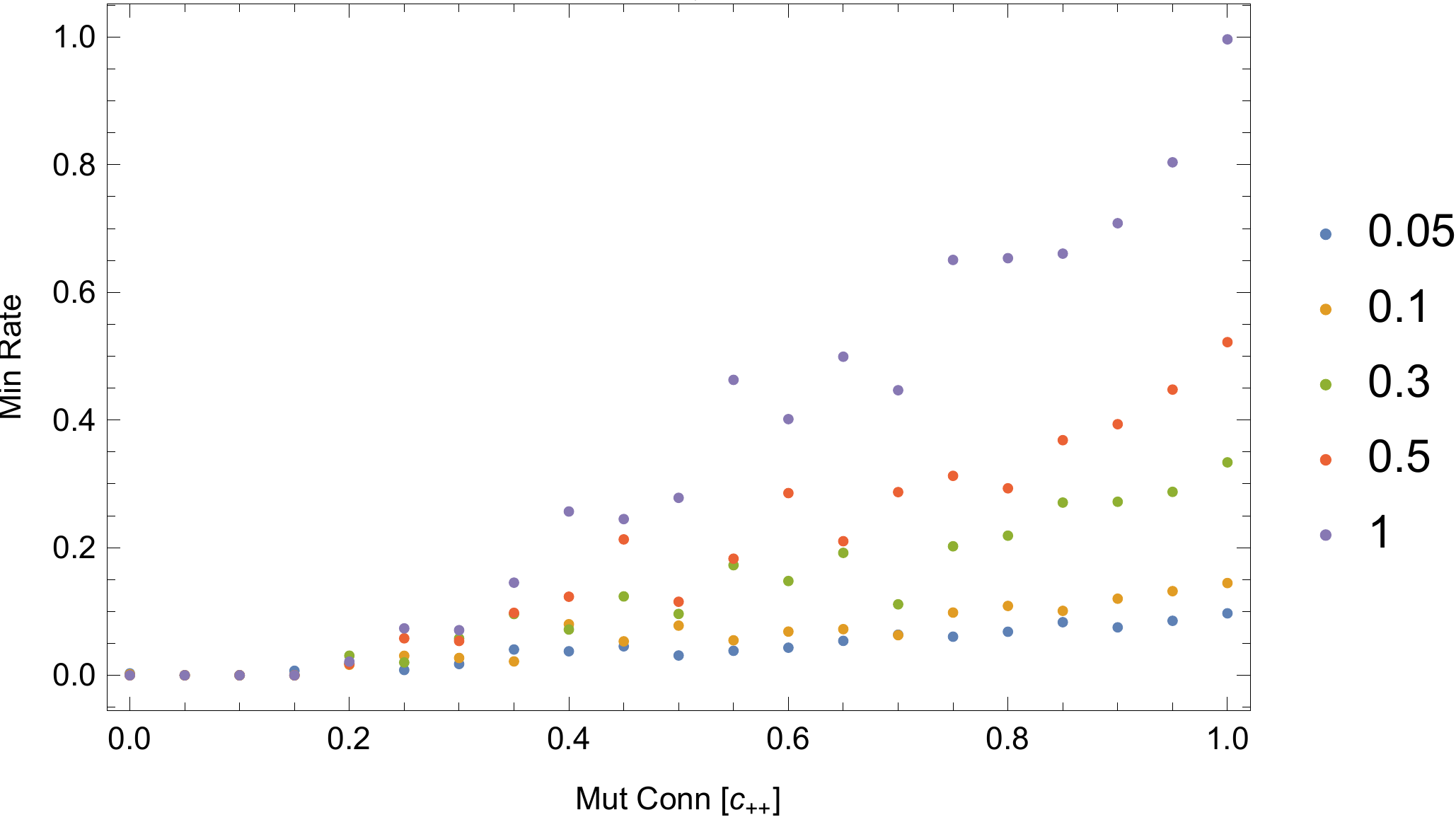}
	\caption{Plot of the Min Rate defined as $\min_{j=1,\dots,S, t\geq 0}\omega(j,\eta(t),M,L)$, where the rates $\omega$ are given by Eq. (1) in the main text and $\eta(t)$ is the mean field solution of Eq. \ref{dlmfL}, as a function of the connectivity of mutualistic $C_M=C_{++}$ and exploitative $C_L=1-C_M$ interactions for different average interaction strengths (colored points) $\mu=\mu_M=\mu_L=0.05,0.1,0.3,1$ (see legend) and $\epsilon_1=\epsilon_2=1$. In all cases the distribution from which interaction strengths are drawn as explained in main text from a bivariate Gaussian distribution with mean $\mu$ and standard deviation ($\sigma=0.01\mu$). The network size considered here is $S=20$. Similar results are found also for $S=50$ and $S=100$. The only cases where the birth rates (given by Eq. 1 in the main text) become negative during the mean field evolution, occur when exploitative interactions are dominant (region for $C_M<0.2$, $C_L > 0.8$).}
	\label{FigRates}
\end{figure}

\begin{figure}[h!]
	\centering
	\includegraphics[width=0.95\columnwidth]{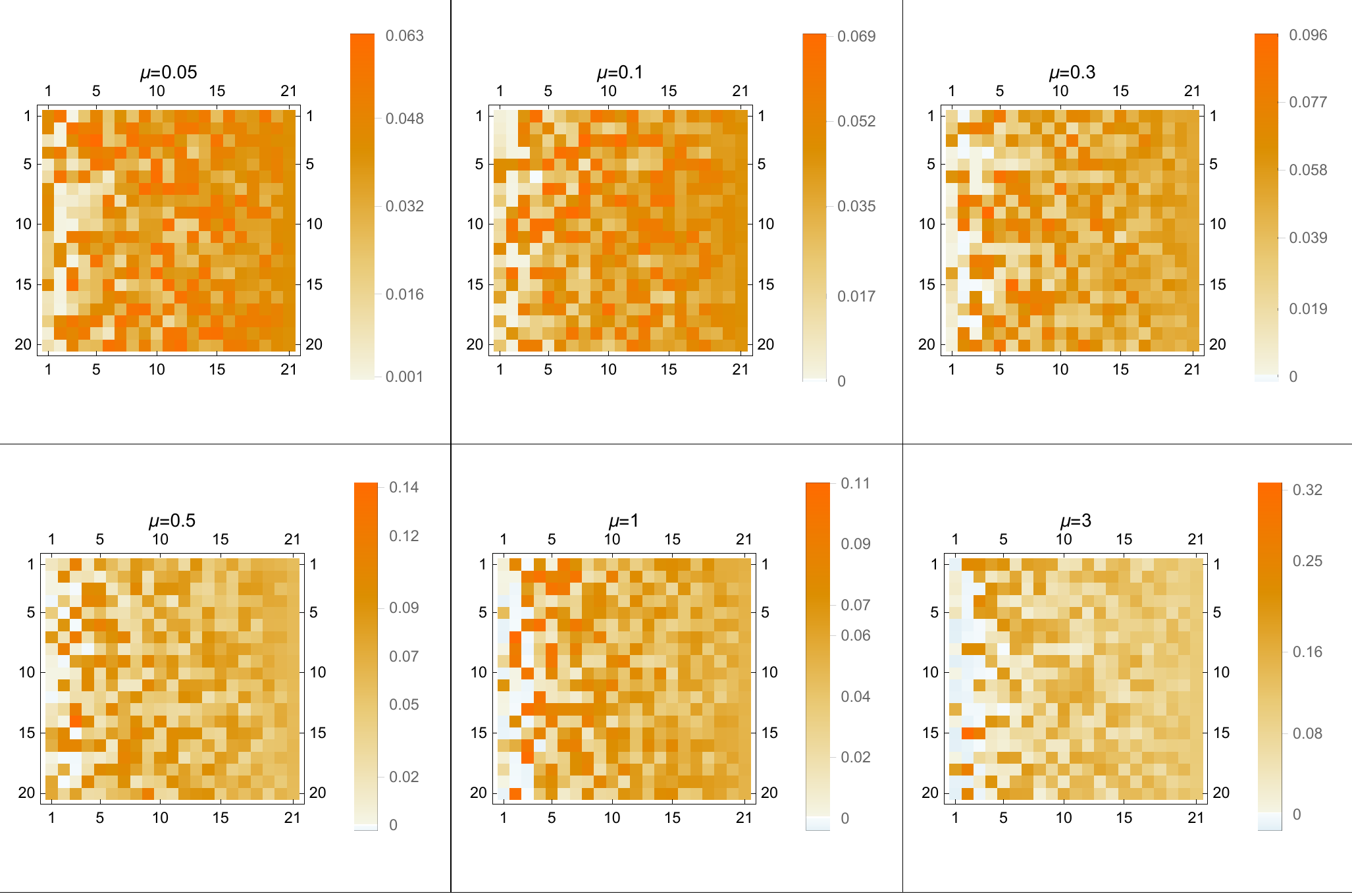}
	\caption{Populations of the species at the stationary state of the dynamics given by Eq. \ref{dlmfL} and the same parameters set in Fig. \ref{FigRates}. The $y$-axis denotes the species label (from $1$ to $20$), while the $21$ points in the $x$-axis represents the $21$ different connectivity configurations: from $C_M=0$ to $C_M=1$ with steps of $\Delta C_M=0.05$ and $C_L=1-C_M$. We numerically checked that as long as the birth rates (given by Eq. 1 in the main text) are positive, then no extinctions are observed (all species populations greater than zero).}
	\label{Populations}
\end{figure}

We have numerically and systematically investigated the number of extinctions in ecological systems with both mutualistic and exploitative species interactions, as a function of different parameters: the average interaction strengths $\mu=\mu_L=\mu_M$, the connectance $C_M$, $C_L$, the network size $S$, etc. 
In all these cases we found that, as long as the birth rates (given by Eq. 1 in the main text) remain positive during the evolution, extinctions are not observed (see Fig. \ref{FigRates} and \ref{Populations}).

\

\subsection*{Stability of the equilibria when $\epsilon_2 \neq 0$}

\begin{figure}[h!]
	\centering
	\includegraphics[width=0.9\columnwidth]{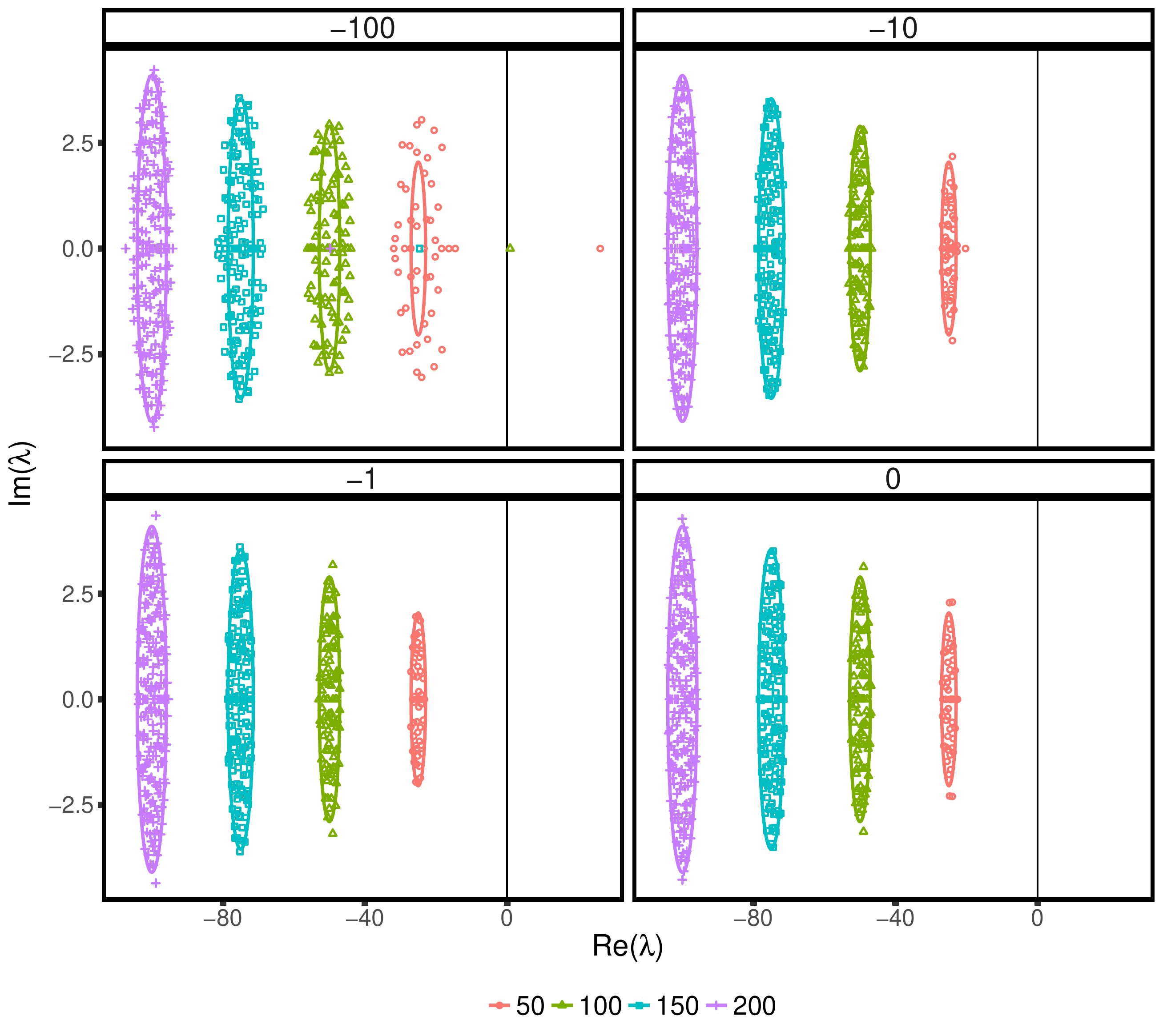}
	\caption{Spectrum of the Jacobian matrix. Different panels correspond to different values of $\epsilon_2 = -100,-10,-1,0$ (as denoted at the top of each inset), while $\epsilon_1 = 1$ for all the simulations. The points are the eigenvalues of one Jacobian matrix obtained sampling at random the matrices $M$ and $L$, whose off-diagonal elements are both drawn uniformly between $0$ and $1$, while the lines indicate the analytical prediction obtained in the Methods section of the main text, in the case $\epsilon_2 = 0$. Colors and shapes correspond to different number of species ($S = 50,100,150,200$ as denoted by the bottom legend). In all the cases, larger matrices turn out to be more stable. The black vertical line indicates the stability threshold.}
	\label{FIGstabLM1}
\end{figure}

\begin{figure}[h!]
	\centering
	\includegraphics[width=0.9\columnwidth]{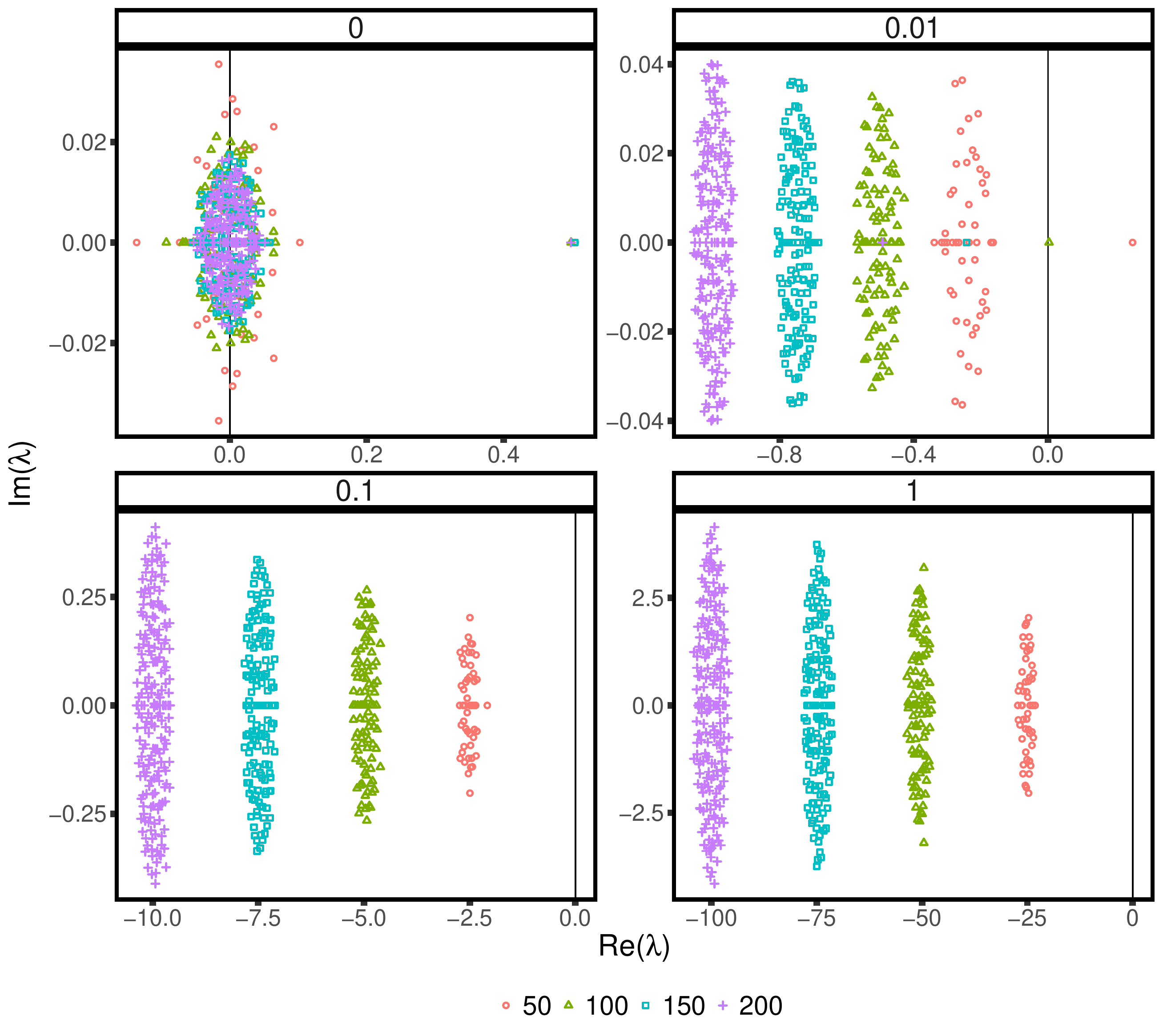}
	\caption{Same as in Fig. \ref{FIGstabLM1} but with $\epsilon_2 = 1$ and varying $\epsilon_1=0,0.01,0.1,1$ (as denoted at the top of each inset).
		Colors and shapes correspond to different number of species. When $\epsilon_1 = 0$, the system is always unstable. As soon as a $\epsilon_1 > 0$ is considered, the spectrum shift on the left, making the system stable. It is important to observe that this happens even for very small values of $\epsilon_1$. The minimum $\epsilon_1$ needed to stabilize the system is in fact expected to go to zero as the number of species $S$ increases ($S = 50,100,150,200$ as denoted by the bottom legend). The off-diagonal elements of the matrices $M$ and $L$ are both drawn uniformly between $0$ and $1$.}
	\label{FIGstabLM2}
\end{figure}

As shown in the Methods section of the main manuscript, the exploitative interactions do not contribute to the stability of the fixed point in the large $S$ limit if $\epsilon_1>0$ (see Fig. \ref{FIGstabLM1} and \ref{FIGstabLM2}). Here we present numerical simulations visualizing this result.

\clearpage


\noindent \textbf{\textit{Author contributions} }
CT and SS contributed equally to this work. SS, MF and AM designed the study, CT, JG and SS performed numerical simulations, AM, MF and JG performed analytical calculations. All authors contributed in writing the manuscript.

\noindent \textbf{\textit{Author declaration} }
The authors declare that they have no competing financial interests.

\noindent \textbf{\textit{Acknowledge} }
We acknowledge enlightening discussions with Stefano Allesina, Sandro Azaele, Jayanth Banavar and Miguel Mu\~noz. S.S., C.T., A.M. acknowledge Fondazione Cariparo for financial support. S.S. acknowledge the Department of Physics and Astronomy, UNIPD for support to SID grant 2017 and the University of Padova for the STARS grant 2017. M.F. was partially supported by the INdAM -- GNAMPA Project 2017 ``Collective periodic behavior in interacting particle systems'' and  by Grant P201/12/2613 of the Czech Science Foundation (GACR).

\clearpage


\bibliography{Ref}
\bibliographystyle{plain}

\end{document}